\documentclass[journal]{IEEEtran}
\usepackage{amssymb}
\usepackage{amsmath, amsthm, amsfonts}

\usepackage{float}

\usepackage{graphicx}
\usepackage{epstopdf}
\AppendGraphicsExtensions{.tif}

\usepackage[utf8]{inputenc}
\usepackage[T1]{fontenc}
\usepackage{lmodern}
\usepackage{amsmath}
\usepackage{mathtools}
\usepackage[ruled, vlined, linesnumbered]{algorithm2e}
\usepackage{bm}

\usepackage{subfig}
\usepackage{ragged2e}

\usepackage[table]{xcolor}
\usepackage{booktabs}
\usepackage{graphicx}
\usepackage{caption}
\usepackage{threeparttable}
\usepackage{tabularx}
\usepackage{multirow}
\usepackage{array}

\usepackage{subfig}
\usepackage{graphicx}

\usepackage{textcomp}
\usepackage{siunitx}

\usepackage{subfig}
\usepackage{ragged2e}

\newcommand{\Tau}{\mathrm{T}}
\begin{document}
\bstctlcite{IEEEexample:BSTcontrol}
    \title{Deep Generative Graph Distribution Learning \\ for Synthetic Power Grids}

\author{Mahdi~Khodayar,~\IEEEmembership{Student Member,~IEEE,}
	Jianhui~Wang,~\IEEEmembership{Senior Member,~IEEE,}
	and~Zhaoyu~Wang,~\IEEEmembership{Member,~IEEE}
	\thanks{Mahdi Khodayar and Jianhui Wang are with the Department
		of Electrical and Computer Engineering, Southern Methodist University, Dallas,
		TX, USA. e-mail: Mahdik@smu.edu and Jianhui@smu.edu. Zhaoyu Wang is with the Department of Electrical and Computer Engineering, Iowa State University, IA, USA. e-mail: wzy@iastate.edu.}
	}

\markboth{
}{Mahdi Khodayar and Jianhui Wang}

\maketitle
\begin{abstract}
Power system studies require the topological structures of real-world power networks; however, such data is confidential due to important security concerns. Thus, power grid synthesis (PGS), i.e., creating realistic power grids that imitate actual power networks, has gained significant attention. In this letter, we cast PGS into a graph distribution learning (GDL) problem where the probability distribution functions (PDFs) of the nodes (buses) and edges (lines) are captured. A novel deep GDL (DeepGDL) model is proposed to learn the topological patterns of buses/lines with their physical features (e.g., power injection and line impedance). Having a deep nonlinear recurrent structure, DeepGDL understands complex nonlinear topological properties and captures the graph PDF. Sampling from the obtained PDF, we are able to create a large set of realistic networks that all resemble the original power grid. Simulation results show the significant accuracy of our created synthetic power grids in terms of various topological metrics and power flow measurements.

\end{abstract}


\IEEEpeerreviewmaketitle

\section{ INTRODUCTION}

\IEEEPARstart{M}{ANY} power system studies require the actual topologies of real-world power networks with real bus and line coordinates as well as real physical charactristics such as power injections and line impedance values. However, due to the confidentiality concerns, there are very few datasets of real power grids such as the IEEE test cases \cite{IEEEtestcase} and the Polish grid \cite{PolishGrid} that are publicly available for such studies. Even these datasets do not contain important information regarding the geographical locations of buses, and lack the diverse characteristics of real topologies. In recent studies, several works are focused on the spatial models for power networks that are realistic but synthesized without revealing any confidential information. \cite{gegner2016methodology, birchfield2017grid} create power networks based on the locations of cities and power plants in Texas; however, no geographical or performance comparison with the actual power grid is provided to justify the approach. Although \cite{gegner2016methodology, birchfield2017grid} contain useful engineering details, they do not devise a general algorithm that can work with any realistic power network; thus, in practice, these approaches are limited. In this line of research, \cite{soltan2018learning} is the only learning-based power grid synthesis (PGS) model presented in recent literature that applies a Gaussian Mixture Model (GMM) to compute the distributions of bus locations in the Western Interconnect network. However, due to lack of generalization, GMM is theoretically incapable of learning the physical properties of power grid components. Moreover, the data required by \cite{soltan2018learning} exceeds the power grid data as this model represents the demand based on the average residential power usage and city populations. Also, similar to \cite{gegner2016methodology} and \cite{ birchfield2017grid}, the GMM approach \cite{soltan2018learning} puts strict  assumptions on the node degree and line distributions which do not necessarily hold. 

This letter presents a novel deep graph distribution learning (DeepGDL) algorithm to create synthetic power grids. We present a recurrent model that efficiently captures node/edge distributions of arbitrary complex \linebreak networks. Leveraging a deep architecture, DeepGDL captures complex nonlinear manifolds of nodes/edges in the actual power network, and generates synthetic power grids that imitate the original network. To the best of our knowledge, DeepGDL is the only deep learning-based PGS solution. Also, in contrast to all previous works, the presented algorithm needs no additional information other than the network topology and the bus/line physical properties; hence, it can be applied as a general framework to any PGS problem with minimum amounts of data.  

\section{DeepGDL for Power Grid Synthesis}
Let us consider the actual power grid obtained from the Columbia University Synthetic Power Grid (CUSPG) dataset \cite{soltan2018columbia} that contains 14430 buses and 18884 lines located in Western US. We represent this network as an undirected weighted graph $G_P = (V,E)$ where $n_i \in V$ denotes the $i$-th node while an edge $l_{i,j} \in E$ is a transmission line connecting two nodes $n_i$ and $n_j$. For each node $n_i$, $G_P$ contains four features including the latitude $x(n_i)$, longitude $y(n_i)$, power supply $pw(n_i)$ (i.e., power generated at the bus $i$), and power demand $pd(n_i)$. A weight $W_{i,j}= X(l_{i,j}) \in \mathbb{R}$ is defined to represent the reactance of the line corresponding to each $l_{i,j}$. Our objective is to learn the PDFs of nodes $n_i \in V$ as well as edges $l_{i,j} \in E$, and sample from them to generate synthetic power grids.

Algorithm 1 is the pseudocode of our presented model. First, using the modularity optimization in \cite{lu2018adaptive}, $G_P$ is decomposed into $K$ communities (i.e., subgraphs) $\Psi=\{C^1, C^2, ..., C^K\}$ where the nodes inside each $C^k ~ (1 \le k \le K)$ are densely connected while the nodes in distinct communities have sparse connections. For each $C^k=(V^k, E^k)$, the adjacency matrix $A^k$ is modeled based on the node ordering $\tau$ that maps any $n_i \in V^k$ to a row/column index $\tau(i)$ in $A^k$. Having the set of all $|V^k|!$ node permutations denoted by $\Tau$, one can write a new adjacency matrix $A^{\tau} \in A^{\Tau}=\{A^\tau | \tau \in \Tau\}$ defined by $A^{\tau}_{i,j} = A^{k}_{\tau(i),\tau(t)}=X(e_{\tau(i),\tau(t)})$; hence, we model $C^k$ as an arbitrarily-sized sequence of node orderings each with a unique nodal and edge feature sequence. Considering the ordering $\tau$, a nonlinear function $f_{n}$ maps $(C^k, \tau)$ to a 2-dimensional node feature tensor $F_{\tau}^k$ using the following formulation:
\begin{equation}\label{eq1}
F_{\tau}^k = f_n(C^k, \tau) = \big(F_{\tau}^k(1),F_{\tau}^k(2),...,F_{\tau}^k(|V^k|)\big) 
\end{equation}
where $F_{\tau}^k(i)=<x(n_{\tau(i)}),y(n_{\tau(i)}), pw(n_{\tau(i)}),pd(n_{\tau(i)})>$ is the nodal feature of the $i$-th node w.r.t the ordering $\tau$. Similarly, we define $f_\Omega$ that maps $(G,\tau)$ to a 2-dimensional edge feature tensor $\Omega_{\tau}^k$ computed by:
\begin{equation} \label{eq2}
\Omega_{\tau}^k = f_\Omega(G,\tau)=\big(\Omega_{\tau}^k(1),\Omega_{\tau}^k(2),..,\Omega_{\tau}^k(|V^k|) \big)
\end{equation}
where $\Omega_{\tau}^k(i)=<A^{\tau}_{1,i},A^{\tau}_{2,i},...,A^{\tau}_{i-1,i}>^T$ is the weighted adjacency vector of node $\tau(i)$; thus, for an undirected weighted $C^k$, the tensor $<F_{\tau}^k,\Omega_{\tau}^k>$ uniquely defines the graph $C^k$. As a result, one can recover $C^k$ by the mapping $F_r(<F_{\tau}^k,\Omega_{\tau}^k>)=C^k$; hence, in order to learn the distribution of the nodes and edges in communities $C^k ~ (k \le K)$ denoted by $P(C^k)$, we maximize the likelihood of their observation in actual network $G_P$ by:
\begin{equation} \label{eq:3}
\displaystyle \small{Max} ~ \sum\limits_{k=1}^{K} [P(C^k) = \sum\limits_{\mathclap{<F_{\tau},\Omega_{\tau}>}}P(F_{\tau}^k,\Omega_{\tau}^k) \mathbb{I}(f_r(F_{\tau}^k,\Omega_{\tau}^k)=C^k)]
\end{equation}
where $\mathbb{I}(x)=1$ if $x$ is true and zero otherwise. Using this optimization, we maximize the observation of $G_P$'s communities, hence finding a model that can imitate $G_P$.

\setlength{\textfloatsep}{0pt}
\begin{algorithm}
\DontPrintSemicolon
\KwIn{Actual Power Network $G_P=(V,E)$}
{
  \textbf{Define} GRU's starting token $START$ and ending token $END$. Community counter $k=1$ 
  \While{$k \le K$}
  {
		Initialize: $[\Omega _\tau^k(0) ,F_\tau^k(0) ] = START$, $i=1$\\
			\While{$[\Omega _\tau^k(i) ,F_\tau^k(i)] \neq END$}
 	    	{	     
      		Compute graph state ${S_i}$ using (\ref{eq:5})\\	
      		Create a new node (i-th node) using (\ref{eq:6})\\
      		$i=i+1$
    		}
    		Generate Synthetic Community $C^k_{Syn}$ using (\ref{eq:6})
            \justify
    		Train GRU, $f_F^{nn}$, and $f_\Omega ^{nn}$ using gradient descent 	
  }
  $P_{Syn} = (V_{Syn}=\cup_{k=1}^{K} V^k_{Syn}, E_{Syn}\cup_{k=1}^{K} E^k_{Syn})$
   \While{$P_{Syn}$ has multiple components}
  {
  \justify
  compute ${\varphi(n_i)}$ as the mean Euclidean distance between $n_i$ and its $10$ closest neighbor nodes in $V_{Syn}$.
  
  \justify
  Select $n_i$ whose degree is at most 2 from all nodes in $V_{Syn}$ using the probability ${\varphi(n_i)}^{-1}$.
  	
\justify
    Add an edge between two nodes $n_i$ and $n_j$ with probability $d(n_j)(Euc_{i,j})^{-1}$. Here, $d(n_j)$ denotes the $n_j$'s degree while $Euc_{i,j}$ is the Euclidean distance of $n_i$ and $n_j$.
  }
  \KwOut{Synthetic Network $P_{Syn} = (V_{Syn},E_{Syn})$}
}
\caption{Deep Graph Distribution Learning}\label{Algorithm3}
\end{algorithm}

Let us write a decomposition for $P({F_\tau^k },{\Omega_\tau^k })$ in (\ref{eq:3}) using conditional probabilities of the observed sequence of nodal and edge features implied by $\tau$ for any community $C^k$:
\begin{equation} \label{eq:4}
\displaystyle \small{ \prod_{i=1}^{|V^k|}{P(} (F_\tau^k(i) ,\Omega_\tau^k(i) )|(F_\tau^k(1) ,\Omega_\tau^k(1))
\\ ,...,(F_{\tau}^{k}(i-1) ,\Omega_{\tau}^{k}(i-1) ))} 
\end{equation}
The recursive characteristics of the graph features in \linebreak (\ref{eq:4}) motivates us to model nodal and edge features by a recurrent neural network. Therefore, we model $P({F_\tau^k },{\Omega_\tau^k })$ using a deep Gated Recurrent Unit (GRU) network. For each community $C^k$, at each iteration $i$ of our deep recurrent model, the GRU computes a graph state $S_i$ using the previous node/edge observations $I(t)$ $(j=1,2,...,i)$ by the following recurrent structure:
\begin{equation} \label{eq:5} 
\begin{array}{l}
\small{{I(t)} = [\alpha^k(t) = F_\tau^k(t),\beta^k(t)=\Omega_\tau^k (i,j)]}\\
\small{z(t)} = Sig ({W_1} \cdot [{I(t)}, {M(t)}])\\ {r(t)} = Sig {({W_2} \cdot [{I(t)},{M(t)}])}\\
\small{{{\tilde M}(t)} = tnh ({W_M} \cdot [{I(t)}, {r(t)} * {M(t-1)}])}\\
\small{{M(t)} = (1 - {z(t)}) * M(t-1)\, + \,{z(t)}\, * \,{{\tilde M}}(t)}
\end{array}
\end{equation}
Here, $\alpha^k(t)$ and $\beta^k(t)$ in $I(t)$ are the intermediate nodal and edge features of $C^k$ observed at the $j$-th round of the recursive formulation. At each round $j$, the temporal features, $z(t)$ and $r(t)$, are computed for the graph observation $I(t)$ using a sigmoid function $Sig$ with weights $W_1$ and $W_2$, respectively. Then, the GRU's updating vector ${{\tilde M}(t)}$ is computed using the intermediate graph features, $r(t)$ and $I(t)$, as well as the GRU state at the previous round $M(t-1)$. Here, $tnh$ denotes a tangent hyperbolic function  with weight $W_M$ applied to compute ${{\tilde M}(t)}$. The GRU state $M(t)$ is finally computed as a linear regression of $M(t-1)$ and ${{\tilde M}(t)}$ weighted by the temporal feature $z(t)$. After running the recursion (\ref{eq:5}) for $i$ rounds, the final GRU state $C(j=i)$ is our extracted graph state $S_i$. 

At each iteration $i$, we model the node/edge distribution of each community $C^k$, and generate a new synthetic community $C^k_{Syn}=(V^k_{Syn}, E^k_{Syn})$ corresponding to $C^k$ using the following formulations:
\begin{equation} \label{eq:6} 
\begin{array}{l}
\xi _i^F = f_F^{nn}({S_i}), \xi _i^\Omega  = f_\Omega ^{nn}({S_i})\\
F_\tau^k(i) \sim {P_{{\xi _i^F}}}\,\,\,\,\,\,\{ Sampling ~ the ~ features ~ of~i- th~node\} \\
\Omega _\tau^k(i) \sim {P_{{\xi _i^\Omega }}}\,\,\,\,\,\,\{ Sampling ~ the ~ weight~of~i- th~edge\} 
\end{array}
\end{equation}

\begin{figure*}[]  
\centering
	\includegraphics[height=1.71in]{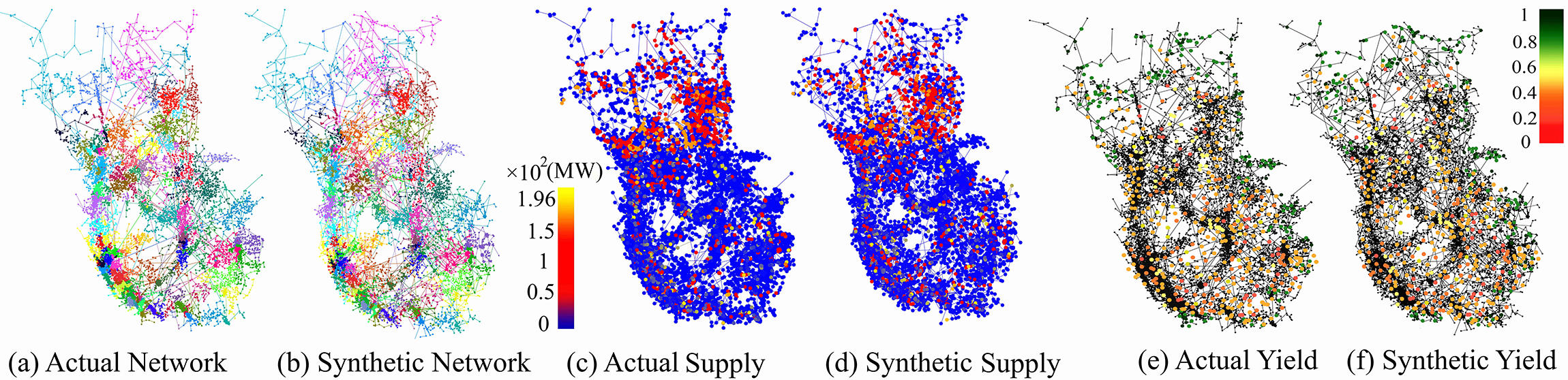}
	\caption{{Visualization of the actual power grid $G_P$ and the synthetic network $G_{Syn}$}}
\label{fig:fig1}	
\end{figure*}
\setlength{\textfloatsep}{0pt}
\begin{table*}[]
\setlength\tabcolsep{1pt} 
\centering
\caption{Topological properties and power flow analysis of the actual and synthetic networks \label{MyTable2}}
\begin{tabular}{|c|c|c|c|c|c|c|c|c|c|c|c|c|c|}
\hline
\rowcolor[HTML]{EFEFEF} 
\cellcolor[HTML]{FFFFFF} & \multicolumn{8}{c|}{\cellcolor[HTML]{EFEFEF}\textbf{Topological Properties}} & \multicolumn{4}{c|}{\cellcolor[HTML]{EFEFEF}\textbf{Power Flow Statistics (MW)}} \\ \hline
\rowcolor[HTML]{EFEFEF} 
\cellcolor[HTML]{FFFFFF}{\color[HTML]{000000} \textbf{\begin{tabular}[c]{@{}c@{}}Networks\end{tabular}}} & \textbf{\#Nodes} & \textbf{\#Edges} & \textbf{$d_{avg}$} & \textbf{$\omega$} & \textbf{$D$} & \textbf{$\rho*10^{-4}$} & \textbf{$Q$} & \textbf{$C$} & {\color[HTML]{000000} \textit{\textbf{\begin{tabular}[c]{@{}c@{}}Average\end{tabular}}}} & {\color[HTML]{000000} \textit{\textbf{\begin{tabular}[c]{@{}c@{}}Median\end{tabular}}}} & {\color[HTML]{000000} \textit{\textbf{\begin{tabular}[c]{@{}c@{}}STDV\end{tabular}}}} & {\color[HTML]{000000} \textit{\textbf{\begin{tabular}[c]{@{}c@{}}Max\end{tabular}}}}  \\ \hline
\rowcolor[HTML]{FFFFFF} 
{\color[HTML]{000000} $G_P$} & \textbf{14430} & \textbf{18554} & \textbf{2.572} & \textbf{15.061} & \textbf{37} & \textbf{1.782} & \textbf{0.953} & \textbf{0.072} & {\color[HTML]{000000} \textbf{168.93}} & {\color[HTML]{000000} \textbf{33.32}} & {\color[HTML]{000000} \textbf{320.93}} & {\color[HTML]{000000} \textbf{4,777.07}}  \\ \hline
\rowcolor[HTML]{FFFFFF} 
{\color[HTML]{000000} $G_{Syn}$} & 14408 & 18302 & 2.510 & 15.253 & 37 & 1.820 & 0.961 & 0.072 & {\color[HTML]{000000} 173.27} & {\color[HTML]{000000} 29.31} & {\color[HTML]{000000} 302.21} & {\color[HTML]{000000} 4,687.83}  \\ \hline
\rowcolor[HTML]{FFFFFF} 
{\color[HTML]{000000} GMM} & 14269 & 19881 & 2.269 & 16.229 & 35 & 1.953 & 0.804 & 0.063 & {\color[HTML]{000000} 110.25} & {\color[HTML]{000000} 42.78} & {\color[HTML]{000000} 479.08} & {\color[HTML]{000000} 5,521,37}  \\ \hline
\rowcolor[HTML]{FFFFFF} 
{\color[HTML]{000000} \textbf{\begin{tabular}[c]{@{}c@{}}DeepDGL\\ Mean(STDV)\end{tabular}}} & \textbf{\begin{tabular}[c]{@{}c@{}}14301\\ ({$\bm{3.1*10^3}$})\end{tabular}} & \textbf{\begin{tabular}[c]{@{}c@{}}17869\\ ({$\bm{4.3*10^3}$})\end{tabular}} & \textbf{\begin{tabular}[c]{@{}c@{}}2.561\\ (0.08)\end{tabular}} & \textbf{\begin{tabular}[c]{@{}c@{}}15.226\\ (0.29)\end{tabular}} & \textbf{\begin{tabular}[c]{@{}c@{}}36.71\\ (0.13)\end{tabular}} & \textbf{\begin{tabular}[c]{@{}c@{}}1.726\\ (0.15)\end{tabular}} & \textbf{\begin{tabular}[c]{@{}c@{}}0.952\\ (0.02)\end{tabular}} & \textbf{\begin{tabular}[c]{@{}c@{}}0.071\\ (0.004)\end{tabular}} & {\color[HTML]{000000} \textbf{\begin{tabular}[c]{@{}c@{}}171.18\\ (11.34)\end{tabular}}} & {\color[HTML]{000000} \textbf{\begin{tabular}[c]{@{}c@{}}30.42\\ (1.20)\end{tabular}}} & {\color[HTML]{000000} \textbf{\begin{tabular}[c]{@{}c@{}}307.18\\ (15.59)\end{tabular}}} & {\color[HTML]{000000} \textbf{\begin{tabular}[c]{@{}c@{}}4,590.25\\ (217.01)\end{tabular}}}  \\ \hline
\end{tabular}
\end{table*} 
\noindent where $f_F^{nn}$ and $f_\Omega ^{nn}$ are two sigmoidal neural networks (SNNs) with the input $S_i$ and output vectors $\xi _i^F$ and $\xi _i^\Omega$, respectively. The vector $\xi _i^F$ encodes the nodal feature distribution of the $i$-th node in $C^k_{Syn}$. Similarly, $\xi _i^\Omega$ encodes the edge weight distribution corresponding to this node. Sampling from these two distributions in (\ref{eq:6}), we create a new node (i.e., the i-th synthesized node) of $C^k_{Syn}$. We run (\ref{eq:6}) to create new nodes and edges until the GRU reaches a predefined $END$ state, which terminates the graph generation process and outputs $C^k_{Syn}$ for every community $C^k$. Each of the generated communities $C^k_{Syn}$ is a component of our final synthetic power network $P_{Syn}$. 

As shown in \cite{soltan2018learning}, in dense regions of real-world power grids, the low-degree buses are generally connected to high-degree buses. Thus, in order to merge the generated components and form a single-component robust network $P_{Syn}$, we connect the low-degree nodes $n_i \in \cup_{k=1}^{K} V^k_{Syn}$ in dense regions, to high-degree nodes $n_j\in \cup_{k=1}^{K} V^k_{Syn}$ until the $P_{Syn}$ becomes single-component. The resulting $P_{Syn}$ is finally reported as the output.

\section{Simulation Results}
First, we set several hyperparameters defined in the DeepGDL algorithm. The GRU state variables $S$, $z$ and $r$, are considered to have $120$ dimensions. Both $f_{f}^{nn}$ and $f_{\Lambda}^{nn}$ are modeled as a SNN with $45$ neurons in its first layer and $35$ neurons in the second layer. The gradient descent is considered to have a learning rate of $10^{-3}$ while its batch size is $30$. In this study, $K=72$ graph communities are considered for $G_P$. We generate $10^2$ synthetic power grids via DeepGDL. Fig. \ref{fig:fig1}(a) and (b) depict the actual network $G_P$ and one of our synthetic networks $G_{Syn}$, respectively. Each community is shown by a distinct color. Moreover, Fig. \ref{fig:fig1}(c) and (d) show the amount of power supply at each bus (node) in the two networks. To compare the performance of the generated grids with the actual grid, we conduct cascading failure experiments using DC power flow. Fig. \ref{fig:fig1}(e) and (f) demonstrate the Yield (i.e. ratio of demand supplied after cascades to the actual value of demand) in the two power grids for cascades initiatd in $10^3$ regions of radius $100$km. Each point shows the computed Yield at the center of its corresponding region. As shown in Fig. \ref{fig:fig1}, $G_{Syn}$ accurately resembles the actual grid regarding the topological characteristics and power flow measurements. Table I reports detailed topological measurements and power flow statistics obtained by $G_P$, DeepGDL algorithm, and the state-of-the-art GMM \cite{soltan2018learning} model. It is shown that DeepGDL has significant improvements over GMM in terms of average node degree $d_{avg}$, average path length $\omega$, network Diameter $D$, Density $\rho$, Modularity $Q$, and Average Clustering Coefficient $C$. Also, Table I shows the substantial accuracy improvements of DeepGDL over the state-of-the-art in terms of power flow measurements.     

\section{Conclusions}
This letter presents a deep graph distribution learning algorithm for the problem of power grid synthesis. A novel recurrent model is proposed to capture deep nodal and edge features from a realistic power network. To the best of our knowledge, DeepGDL is the only deep learning model that can synthesize large-scale networks. Simulation results show significant accuracy of DeepGDL in terms of topological measurements and power flow analysis metrics.

\bibliographystyle{IEEEtran}
\bibliography{IEEEabrv,Bibliography}

\begin{thebibliography}{1}
\providecommand{\url}[1]{#1}
\csname url@rmstyle\endcsname
\providecommand{\newblock}{\relax}
\providecommand{\bibinfo}[2]{#2}
\providecommand\BIBentrySTDinterwordspacing{\spaceskip=0pt\relax}
\providecommand\BIBentryALTinterwordstretchfactor{4}
\providecommand\BIBentryALTinterwordspacing{\spaceskip=\fontdimen2\font plus
\BIBentryALTinterwordstretchfactor\fontdimen3\font minus
  \fontdimen4\font\relax}
\providecommand\BIBforeignlanguage[2]{{%
\expandafter\ifx\csname l@#1\endcsname\relax
\typeout{** WARNING: IEEEtran.bst: No hyphenation pattern has been}%
\typeout{** loaded for the language `#1'. Using the pattern for}%
\typeout{** the default language instead.}%
\else
\language=\csname l@#1\endcsname
\fi
#2}}

\bibitem{IEEEtestcase}
``{IEEE} benchmark systems,'' \url{www.labs.ece.uw.edu/pstca/}, accessed:
  2019-01-20.

\bibitem{PolishGrid}
``Polish grid,'' \url{www.pserc.cornell.edu/matpower/}, accessed: 2019-01-21.

\bibitem{gegner2016methodology}
K.~M. Gegner, A.~B. Birchfield, T.~Xu, K.~S. Shetye, and T.~J. Overbye, ``A
  methodology for the creation of geographically realistic synthetic power flow
  models,'' in \emph{Power and Energy Conference at Illinois (PECI), 2016
  IEEE}, 2016, pp. 1--6.

\bibitem{birchfield2017grid}
A.~B. Birchfield, T.~Xu, K.~M. Gegner, K.~S. Shetye, and T.~J. Overbye, ``Grid
  structural characteristics as validation criteria for synthetic networks,''
  \emph{IEEE Transactions on power systems}, vol.~32, no.~4, pp. 3258--3265,
  2017.

\bibitem{soltan2018learning}
S.~Soltan, A.~Loh, and G.~Zussman, ``A learning-based method for generating
  synthetic power grids,'' \emph{IEEE Systems Journal}, no.~99, pp. 1--10,
  2018.

\bibitem{soltan2018columbia}
S.~Soltan, A.~Loh, and g.~Zussman, ``Columbia university synthetic power grid
  with geographical coordinates,'' Tech. Rep., 1 2018.

\bibitem{lu2018adaptive}
X.~Lu, K.~Kuzmin, M.~Chen, and B.~K. Szymanski, ``Adaptive modularity
  maximization via edge weighting scheme,'' \emph{Information Sciences}, vol.
  424, pp. 55--68, 2018.

\end{thebibliography}
\let\mybibitem\bibitem

\vfill

\end{document}